%% file: template.tex
\documentclass{Interspeech2024}

\usepackage[printonlyused]{acronym} 
\usepackage{url}
\usepackage{graphicx}
\usepackage[table,xcdraw]{xcolor}
\usepackage{amsmath}
\usepackage{amssymb}
\usepackage{amsfonts}
\usepackage[caption=false]{subfig}
\usepackage{flushend} 
\usepackage[citecolor=black,
            colorlinks=true,
            linkcolor=black,
            urlcolor  = black,
            pdftitle={Transcription-Free Fine-Tuning of Speech Separation Models for Noisy and Reverberant Multi-Speaker Automatic Speech Recognition},
            pdfauthor={William Ravenscroft, George Close, Stefan Goetze, Thomas Hain,
Mohammad Soleymanpour, Anurag Chowdhury, Mark C. Fuhs},
            pdfkeywords={},
            pdfsubject={}]{hyperref}
\usepackage[english]{babel}
\addto\extrasenglish{}
\addto\extrasenglish{}
\addto\extrasenglish{}
\addto\extrasenglish{}
\addto\extrasenglish{}
\addto\extrasenglish{}

\input{preamble}

\title{Transcription-Free Fine-Tuning of Speech Separation Models for Noisy and Reverberant Multi-Speaker Automatic Speech Recognition
\texorpdfstring{\thanks{
This work was supported by the Centre for Doctoral Training in Speech and Language Technologies (SLT) and their Applications funded by UK Research and Innovation [grant number EP/S023062/1]. This work was also supported in part by Solventum and Toshiba Cambridge Research Laboratory.\\
* Work done while an intern at Solventum, formerly part of 3M.
}
\vspace{-0.6cm}
}{}
}

\name[affiliation={1,2*}]{William}{Ravenscroft}
\name[affiliation={1}]{George}{Close}
\name[affiliation={1}]{Stefan}{Goetze}
\name[affiliation={1}]{Thomas}{Hain}
\name[affiliation={2}]{\texorpdfstring{\\Mohammad}{Mohammad}}{Soleymanpour} 
\name[affiliation={2}]{Anurag}{Chowdhury} 
\name[affiliation={2}]{Mark C.}{Fuhs} 

\address{
\texorpdfstring{
  $^1$Department of Computer Science, The University of Sheffield, United Kingdom\\
  $^2$Solventum, USA}{Department of Computer Science, The University of Sheffield, United Kingdom. $^2$Solventum, USA}
}

\email{\texorpdfstring{jwravesncroft1@sheffield.ac.uk\vspace{-0.3cm}}{jwravesncroft1@sheffield.ac.uk}}

\keywords{speech recognition, speech separation, multi-speaker, adaptation, fine-tuning}

\begin{document}

\maketitle


\begin{abstract}
    One solution to \ac{ASR} of overlapping speakers is to separate speech and then perform ASR on the separated signals. Commonly, the separator produces artefacts which often degrade \ac{ASR} performance. Addressing this issue typically requires reference transcriptions to jointly train the separation and \ac{ASR} networks. This is often not viable for training on real-world in-domain audio where reference transcript information is not always available. This paper proposes a \textit{transcription-free} method for joint training using only audio signals. The proposed method uses embedding differences of pre-trained ASR encoders as a loss with a proposed modification to \ac{PIT} called \textit{guided} PIT (GPIT).
    The method achieves a $6.4$\% improvement in \ac{WER} measures over a signal-level loss and also shows enhancement improvements in perceptual measures such as \ac{STOI}.
\end{abstract}

\section{Introduction}
While significant progress has been made in recent years in multi-speaker \ac{ASR} research, it remains a challenging problem \cite{FFASRHaebUmbach,IWAENCbestpaper}. Many different approaches have been proposed to solving this problem \cite{orcwer,sot,vonneumann2023meeting}. 
These methods include entirely end-to-end single models \cite{sot} and modular approaches \cite{chime6,IWAENCbestpaper,vonneumann2023meeting}.

Modular approaches often include a variety of sub-components such as speech separation~\cite{atttasnet}, speaker diarization~\cite{clarke23diarisation} and \ac{ASR} \cite{chime6,vonneumann2023meeting}. Sometimes these sub-components, or \textit{modules}, can be combined in an end-to-end fashion that allows for joint training of the modules \cite{mimospeech,orcwer}. 
More recently, \ac{SOT} was proposed, a technique that enables the design of explicitly multispeaker \ac{ASR} models with a single output sequence containing subsequences separated by a unique token for each speaker \cite{sot}. There is a benefit to having a simplified model that requires no fine-tuning on sub-components but this results in lower interpretability. 
A commonality of both approaches to multi-speaker \ac{ASR} is that they are reliant on the existence of ground truth transcriptions of the speech of each speaker in the mixture. Whilst this is easily obtainable for simulated mixtures, it is often impossible to obtain for `real-world' in-domain speech mixtures. 

In \cite{pandp,mcmganpp}, computing embedding differences between \acp{SSSR} was used as a loss for training speech enhancement networks to improve perceptual speech quality. Furthermore, in \cite{mimic} it was shown that computing the difference between \textit{senone} representations of \ac{ASR} models can be used to adapt speech enhancement networks for robust \ac{ASR}. Recent work~\cite{tu22_interspeech,mogridge2024nonintrusive} has shown that features derived from \ac{ASR} models can capture quality and intelligibility-related information.

In this work, a novel method for transcription-free fine-tuning of a separate-and-recognize modular approach is proposed. 
Embedding differences from a pre-trained \ac{ASR} encoder are used to compute a loss from the reference speech and separated speech signals. Furthermore, a variant of the the standard \ac{PIT}~\cite{pit} algorithm, \acf{GPIT} is proposed to properly address speaker permutation solving of the \ac{ASR} encoder embeddings within the proposed loss function. 
This recognition-based loss term is then used to fine-tune a speech separator.
The proposed method notably improves multi-speaker \ac{WER} performance over traditional signal-level losses across multiple \ac{ASR} models, and also shows improved performance in intrusive perceptual measures. Another benefit of the proposed approach 
is that it allows for fine-tuning separators using truncated audio signals \cite{tsllimits} as the full transcription is not required, potentially resulting in accelerated training, lower memory requirements and reduced computational expenditure.
The proposed method also makes training on real-world data, with pseudo-reference audio
used as input to the proposed loss function, possible~\cite{kalda2024pixit}
.

The remainder of this paper proceeds as follows: in \autoref{tfftm}, the proposed transcription-free fine-tuning method is described in detail. In \autoref{experiments}, the experimental setup and data are described. Results and conclusions are given in \autoref{results} and \autoref{conclusions}, respectively.

\section{Transcription-Free Fine-Tuning Method}\label{tfftm}
In this section, the main components of the proposed transcription-free fine-tuning method are described.
A modularized approach to multi-speaker \ac{ASR} is used; this approach is referred to here as the \textit{separate-and-recognize} approach. 
A noisy reverberant mixture $x[i]$ of $C$ speech signals $s_c[i]$, $c\in\{1,\ldots,C\}$ for time index $i$ is defined as  
\begin{equation}\label{ss:eq:sm:sess}
         x[i] = \sum_{c=1}^C h_{c}[i]\ast s_{c}[i] + \nu[i],
\end{equation}
where $\ast$ denotes the convolution operator, $h_c[i]$ is the room impulse response corresponding to speaker $c$ and $\nu[i]$ denotes additive noise. 
In the separate-and-recognize approach, a separation model firstly separates the mixture signal $x[i]$ in $C$ estimated speech signals $\hat{s}_c[i]$ and each separated signal is fed into an \ac{ASR} module to attain predicted transcription of discrete character tokens $\hat{\vek{t}}_c$ \cite{IWAENCbestpaper}.

\begin{figure}[!t]
    \centering
    \includegraphics[width=0.9\columnwidth]{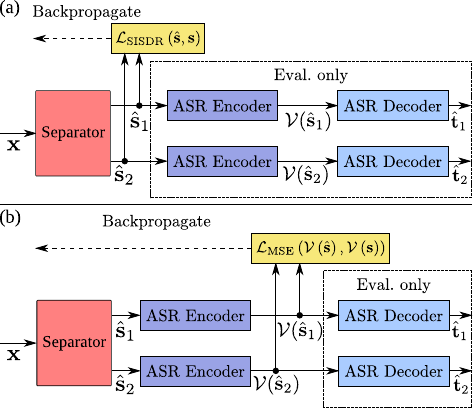}
    \caption{(a) Baseline approach to training speech separators without ASR-based fine-tuning. (b) Proposed fine-tuning approach without using reference transcriptions. Solid lines indicate information flow; dashed lines the direction of gradient backpropagation. Figure exemplary for $C=2$ speakers.}
    \label{fig:aeloss}
\end{figure}

The proposed fine-tuning method assumes the existence of pre-trained speech separation and \ac{ASR} networks, where the speech separation model has been trained using a conventional signal-level objective function such as \ac{SISDR} \cite{tasnet}. In the proposed method, these two modules go through an additional number of \acp{ATE}, whereby the parameters of the \ac{ASR} model are frozen and the embedding differences of the \ac{ASR} encoder are backpropagated through the separation network where the parameters are updated at each step. The proposed approach is shown in \autoref{fig:aeloss} compared to the baseline model trained with a purely signal-level \ac{SISDR} loss.

\subsection{ASR Encoder Loss}
This section describes the proposed \ac{AE} loss used for fine-tuning the speech separation model. In this work the \ac{ASR} encoder used is a \ac{CTC}-based model \cite{ctc} with a discrete number of possible output symbols $N$.
The encoder network is defined as a function
\begin{equation}
    \mathcal{V}:\Real^{L_x}\mapsto\Real^{L \times N}
\end{equation}
 where $L_x$ is the length of an input speech signal $s_c[i]$, $L$ is the output sequence length and $N$ is the number of possible symbols, i.e.~the number of characters that can be interpreted by a decoder or decoding function (minimum of $26$ plus a word boundary token and a blank symbol for the Latin alphabet).

The proposed \ac{AE} loss computes the difference between the encoder output of the predicted speech signals $\hat{s}_c$ and reference speech signal ${s}_c$, $\forall c \in \{1,\ldots, C\}$. The loss is the \ac{MSE} of the two output sequences defined as 
\begin{equation}
    \mathcal{L}_\mathrm{AE}\left(\vek{s}_c,\hat{\vek{s}}_c\right)=\frac{1}{LN} 
    \sum_{\ell=1}^{L} 
    \sum_{w=1}^{N}
    \left(\mathcal{V}(\hat{\vek{s}}_c)_{\ell,n} - \mathcal{V}(\vek{s}_c)_{\ell,n} \right)^2
    \label{eq:ae_loss}
\end{equation}
where ${s}_c$ is the target audio and  $\hat{s}_c$ is the predicted output audio of the seperator for speaker $c$. 
The $\mathcal{L}_\mathrm{AE}$ loss is analogous to a conventional time-frequency based loss, but instead of comparing references and outputs in terms of continuous time/frequency bins, the proposed loss instead compares in terms of discrete time/character label logits. 

\subsection{Guided Permutation Invariant Training (GPIT)}
Empirically it was found that the standard \ac{PIT} algorithm \cite{upit} was not robust enough against distortions in the ASR encoder features when applying the proposed \ac{AE} loss function in (\ref{eq:ae_loss}), resulting in inaccuracies in resolving speaker permutations.
This resulted in models trained using the \ac{AE} loss diverging and giving $100\%$ \ac{WER} in evaluations. 
Thus, a modified scheme referred to as \acf{GPIT} (not to be confused with Graph-PIT \cite{graphpit}) is proposed that uses an alternate signal-level loss to guide the permutation solving, and then the \ac{AE} loss is applied to the minimum permutation estimates and references.
The permutation-solving formula is defined the same as for \ac{PIT}. The predicted permutation $\hat{\phi}$ is determined by 
\begin{equation}
    \hat{\phi} = \argmin_{\phi\in{\vek{\Phi}}}\sum_{c=1}^{C}\mathcal{L}_\mathrm{guide}\left(\hat{\vek{s}}_c^{(\phi)},\vek{s}_c\right)
\end{equation}
where $\phi$ is a permutation in the set $\vek{\Phi}$ of $\Phi=C!$ possible permutations and $\hat{\vek{s}}_c^{(\phi)}$ denotes the $\phi$th possible permutation of $\hat{\vek{s}}_c$. $\mathcal{L}_\mathrm{guide}$ denotes the \textit{permutation guiding} loss. In this paper $\mathcal{L}_\mathrm{guide}=\mathcal{L}_\mathrm{SISDR}$ is chosen. Using the predicted permutation the desired loss, in this case $\mathcal{L}_\mathrm{AE}(\cdot)$ from (\ref{eq:ae_loss}), is calculated s.t.~the final loss value is
\begin{equation}
\mathcal{L}_\mathrm{GPIT}\left(
\vek{s}_c,\hat{\vek{s}}_c, \hat{\phi}
\right)
=
\sum_{c=1}^C\mathcal{L}_\mathrm{AE}\left(\vek{s}_c,\hat{\vek{s}}_c^{(\hat{\phi})}\right).
\end{equation}

\input{tables/main_results}
\section{Experimental Setup}\label{experiments}
\subsection{Data}
The WHAMR \cite{WHAMR} corpus is used for all experiments in this work. This corpus is an extension of the WSJ0-2Mix corpus \cite{Isik} which in turn is an extension of the WSJ0 corpus \cite{wsj0}. WHAMR is a corpus of noisy, reverberant, two-speaker mixtures created from overlapping two artificially reverbed speaker utterances from WSJ0 at \textit{mixing \acp{SNR}} of $0$-$5$~dB and then adding an ambient noise source at an \ac{SNR} of $-6$-$3$~dB relative to the loudest speaker.
The $16$kHz \textit{max} (non-truncated) configuration of the corpus is used. 
Dynamic mixing \cite{wavesplit} is also used, whereby the training set is uniquely simulated at each epoch to improve data diversity and model generalization. Signal lengths are also limited to $4$s in length to speed up training as well as reduce computational expenditure and memory consumption \cite{tsllimits}. Further analysis of \ac{TSL} limits is given in \autoref{results:tsllimits}.

\subsection{Speech Separator}
The TD-Conformer-XL model is chosen for speech separation as it gives close to state-of-the-art performance on the WHAMR benchmark and is open source\footnote{TD-Conformer training recipe available at  \url{https://github.com/jwr1995/PubSep}} \cite{tdConformer}. 
The configuration is based on the best-performing model on WHAMR in \cite{tdConformer} with $S=1$ subsampling layers, a kernel size of $P=64$ and feature dimension $B=1024$.
To compensate for the larger sampling rate used in this paper ($8$kHz~vs~$16$kHz here), the number of subsampling layers is set to $S=2$.
This model is \textit{pre-trained} on $4$s audio clips with a learning rate of $\eta_\mathrm{PT}=5\times 10^{-5}$ over 150 epochs. The learning rate is halved after $90$ epochs if there is no improvement in $3$ epochs.

\subsection{Speech Recognizers}
The large \ac{SSSR}-based Wav2Vec2 model \cite{wav2vec2} fine-tuned on 960 hours of LibriSpeech data with a \ac{CTC} loss is chosen\footnote{Torchaudio Wav2Vec2 Large model available at  \url{https://pytorch.org/audio/0.10.0/pipelines.html\#torchaudio.pipelines.WAV2VEC2_ASR_LARGE_960H}} as the primary model used to obtain the embeddings $\mathcal{V}(\cdot)$ in the \ac{AE} loss in (\ref{eq:ae_loss}), as well as for evaluating \ac{ASR} performance.
The Wav2Vec2 model is a large transformer \ac{SSSR} model that is initially trained to predict speech representations from time-domain signals via self-supervision where the model is trained to predict contextualized representations across masked portions of the input signal from the unmasked signal context \cite{wav2vec2}.
The fine-tuning of Wav2Vec2 involves training an additional linear layer to map these representations to a \ac{CTC} specific representations, in this case, grapheme-based representations where each logit represents a possible grapheme.
Note that the fine-tuning in this model is unrelated to the proposed fine-tuning method described in \autoref{tfftm}.
In the large Wav2Vec2 model used in the following experiments, the number of labels is $N=30$ ($28$ characters interpreted verbatim plus word boundary token ``\textbar" and blank symbol ``-"). 
In the \ac{CTC} decoder of the Wav2Vec2 ASR model, an open-source 4-gram Librispeech language model is used\footnote{Librispeech language model and other resources available at: \url{https://www.openslr.org/11/}.}.

In addition, the large Whisper model\footnote{In this work, Whisper 'large-v2' is used, which can be downloaded from: \url{https://github.com/openai/whisper}.} \cite{whisper} is used as an \textit{unseen} ASR evaluation model for evaluating ASR performance on a model that was not used in the fine-tuning of the separator. Whisper is a weakly supervised speech foundation model that uses multi-task training. The weakly supervised and multi-task approaches are designed to make the ASR model generalisation well across many acoustic conditions and speaker types \cite{whisper}. Thus this well-generalising \ac{ASR} model is thought to make it more challenging for the proposed approach to achieve improvements over the baseline SISDR models.

\subsection{Fine-Tuning}
For fine-tuning with the \ac{AE} loss and \ac{GPIT}, a learning rate of $\eta_{\mathrm{FT}}=2\times 10^{-7}$ is used. Fine-tuning is performed over $30$ \acfp{ATE} with the learning rate fixed for all epochs. Exploring different learning rate strategies and optimizing this hyperparameter is beyond the scope of this paper and left to future work.

\subsection{Evaluation Metrics}
Several options are available for multi-speaker \ac{WER} evaluation \cite{meeteval}. In this paper, two such definitions are chosen: \ac{CP-WER} \cite{chime6} and \ac{ORC-WER} \cite{orcwer}. The key difference between the two measures is that \ac{CP-WER} penalizes output speaker channel switches and \ac{ORC-WER} is unconcerned with whether a given speaker is output on a single channel or multiple channels, so long as the ASR model is still able to estimate the word accurately. \Ac{CP-WER} is thus the more important measure as, ideally, in the proposed system, the goal is to have one speaker for each output channel. However, the addition of \ac{ORC-WER} provides additional insight into the overall intelligibility of the speech regardless of which channel(s) a given speaker gets output to. Intrusive speech enhancement measures are also used to observe any benefit gained in these using the proposed approach. \Ac{STOI} \cite{stoi} is used to measure speech intelligibility, \ac{PESQ} \cite{PESQ} is used to measure speech quality and \ac{SRMR} \cite{srmr} is used to assess reverberant effects.

\section{Results}\label{results}

\subsection{Results on clean targets}
The performance of the \ac{AE} loss function (\ref{eq:ae_loss}) using the large Wav2Vec2 model is shown in \autoref{tab:wav2vec2}.
The TD-Conformer separation network trained with the proposed \ac{AE} loss for $30$ \acp{ATE} is compared to the TD-Conformer separator before fine-tuning and the TD-Conformer trained with an additional $30$ epochs, but using the standard signal-level \ac{SISDR} loss \cite{tdConformer,tasnet}. 
The model trained with the proposed \ac{AE} loss significantly outperforms both models in ASR performance (\ac{CP-WER} and \ac{ORC-WER}). 
The \ac{AE} loss-based model also outperforms the others in terms of the speech enhancement metrics \ac{PESQ}, \ac{STOI} and \ac{SRMR}. This is a powerful finding, as often improved perceptual performance leads to degraded \ac{ASR} performance and vice-versa. It is consistent however with prior work~\cite{pandp} which similarly compares \ac{SSSR} output representations in a speech enhancement system loss function.

\subsection{Generalization to an Unseen ASR System}
\input{tables/whisper_results}
The generalization of the improvements found using the proposed \ac{AE} loss in \autoref{tab:wav2vec2} is analysed in this subsection by reevaluating models on the large Whisper ASR model \cite{whisper} that was not used in the fine-tuning stages. 
The results in \autoref{tab:whisper} show a consistent improvement in both \ac{CP-WER} and \ac{ORC-WER} for the \ac{AE} loss over the SISDR loss trained with additional epochs demonstrating that at least some of the improvements can generalise from one \ac{ASR} model to another.

\subsection{Joint SISDR Loss Weighting}
To further assess the impact of the proposed \ac{AE} loss $\mathcal{L}_\mathrm{AE}$ against the \ac{SISDR} loss $\mathcal{L}_\mathrm{SISDR}$~\cite{tdConformer,tasnet}, a series of experiments fine-tuning the separator with a joint weighted loss of the two different loss types is carried out. 
The joint loss is defined as
\begin{equation}
    \mathcal{L}_\mathrm{Joint} = (1-\alpha) \mathcal{L}_\mathrm{AE} + \alpha\mathcal{L}_\mathrm{SISDR}
    \label{eq:joint_loss}
\end{equation}
where $\alpha$ 
controls the weighting of the two loss terms. Five models are fine-tuned with values of $\alpha\in\{0.0,0.2,0.4,0.6,0.8,1.0\}$ for $30$ \acp{ATE}.
\begin{figure}[!ht]
    \centering
    \includegraphics[trim={0 0 0 0.3cm},clip]{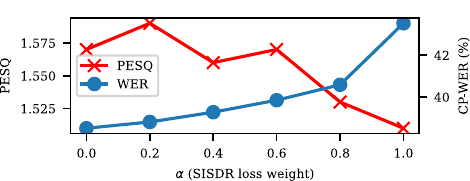}
    \caption{\ac{ASR} performance (CP-WER) and objective perceptual quality of test audio for models trained with differing weight $\alpha$  between loss terms in (\ref{eq:joint_loss}).}
    \label{fig:joint_loss_plot}
\end{figure}

\autoref{fig:joint_loss_plot} shows the performance of the models fine-tuned using (\ref{eq:joint_loss}) in terms of \ac{CP-WER} and \ac{PESQ} on the WHAMR test set. The \ac{ASR} performance of the models trained with lower values of $\alpha$ is significantly better than those trained with higher values,
as the influence of the proposed $\mathcal{L}_\mathrm{AE}$ function on the models' training is higher for lower values of $\alpha$. The biggest jump in improvement is between $\alpha = 1.0$ (i.e no use of $\mathcal{L}_\mathrm{AE}$) and $\alpha = 0.8$, suggesting that even a small weighting of the proposed loss can significantly improve \ac{ASR} performance. 
Perceptual quality in terms of \ac{PESQ} does not change as uniformly or significantly for lower $\alpha$ values, but a trend is still apparent; models with a greater weighting of the proposed loss $\mathcal{L}_\mathrm{AE}$ produce higher quality audio.\\ 

\subsection{Training Signal Length Analysis}\label{results:tsllimits}
A benefit of the transcription-free method is the ability to train on arbitrary signal lengths. In transcription-dependant losses, this is non-trivial due to the requirement of truncating the transcription in alignment with the audio. The impact of applying \ac{TSL} limits \cite{tsllimits,rfield} is analysed in \autoref{tab:tsllimits}.
\input{tables/TSL_table}
The results show that using longer signal lengths gives slightly better performance, most likely due to the ASR Encoder having more context. This is opposite to the SISDR-based evaluations in \cite{tsllimits} where improved SISDR performance could be obtained on WHAMR using shorter signal lengths.

\subsection{ASR Encoder Visualization}
\begin{figure}[!ht]
\centering
\includegraphics[width=0.7\columnwidth]{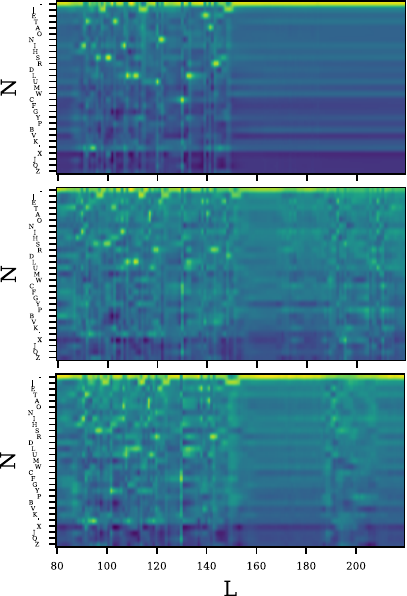}
  

  

  
 \caption{Wav2Vec2 ASR Encoder output representations for reference audio $\mathcal{V}(s[n])$ (top), and $\mathcal{V}(\hat{s}[n])$ for models with baseline $\mathcal{L}_\mathrm{SISDR}$ (middle) and proposed $\mathcal{L}_\mathrm{AE}$ fine-tuning (bottom) losses}\label{fig:specs}
\end{figure}

A visual comparison of Wav2Vec2 \ac{ASR} Encoder output representations $\mathcal{V}(\cdot)$ used in \eqref{eq:ae_loss} of audio outputs from models trained using SISDR loss (middle) and AE loss (bottom) can be seen in \autoref{fig:specs}. The ASR encoder output representation $\mathcal{V}(\cdot)$ of the respective reference audio is visualized in the top panel.
Notably, the \ac{AE} loss reduces the uncertainty in the logits when compared to the SISDR loss. In both representations of the estimated signals, there is some bleed-through of interfering noise and speakers at $L\gtrapprox180$, but this effect is reduced in the \ac{AE} loss suggesting that the additional context provided by the encoder network may improve separation performance and reduce more general distortions.

\section{Conclusions}\label{conclusions}
This paper presented a novel method for fine-tuning speech separation models for multi-speaker \ac{ASR} without the need for reference transcriptions. It was shown that the proposed loss leveraging pre-trained \ac{ASR} encoder representations for fine-tuning separators results in improved \ac{ASR} performance over a standard signal-level-based loss. It was also demonstrated that the proposed method produces improved performance in speech enhancement metrics such as \ac{PESQ} and \ac{STOI} which is often not the case for many other approaches. Finally, it was shown that the performance gained by the proposed method generalises to other \ac{ASR} models that weren't used in the fine-tuning stage.

\bibliographystyle{IEEEtran}
\bibliography{mybib}

\end{document}

%% file: preamble.tex
\newcommand{\vek}[1]{\ensuremath{\mathbf{#1}}}    

\newcommand{\Real}{\mathbb{R}}

\DeclareMathOperator*{\argmin}{arg\,min}

    \acrodef{AC}    {acoustic condition}
    \acrodef{AE}  {ASR encoder}
    \acrodef{ATE}   {addtional training epoch}
    \acrodef{ANN}   {artificial neural network}
    \acrodef{ASR}   {automatic speech recognition}
    \acrodef{BLSTM} {bidirectional long short term memory}
    \acrodef{BS}    {batch size}
    \acrodef{BSS}   {blind source separation}
    \acrodef{CB}    {convolutional block}
    \acrodef{CD}    {convolutional decoder}
    \acrodef{CE}    {convolutional encoder}
    \acrodef{CN}    {channel-wise layer normalization}
    \acrodef{CSM}   {clean speech mixtures}
    \acrodef{cLN}   {cummulative layer normalization}
    \acrodef{Conformer}{convolution augmented Transformer}
    \acrodef{Conv-TasNet}{convolutional time-domain audio separation network}
    \acrodef{ConSepT}{convolutional separation Transformer}
    \acrodef{CP-WER}{concatenated minimum-power word error rate}
    \acrodef{CTC}   {connectionist temporal classification}
    \acrodef{CV}    {computer vision}
    \acrodef{DAE}   {denoising autoencoder}
    \acrodef{DANet} {deep attractor network}
    \acrodef{DC}    {deep clustering}
    \acrodef{DE}    {density estimation}
    \acrodef{DL}    {deep learning}
    \acrodef{DM}    {dynamic mixing}
    \acrodef{DNN}   {deep neural network}
    \acrodef{DPRNN} {dual path recurrent neural network model}
    \acrodef{DPTNet}{dual path Transformer network}
    \acrodef{D-Conv}{depthwise convolution}
    \acrodef{DD-Conv}{deformable depthwise convolution}
    \acrodef{DDS-Conv}{deformable depthwise-separable convolution}
    \acrodef{DP}    {dual-path}
    \acrodef{DS-Conv}{depthwise separable convolution}
    \acrodef{DTCN}  {deformable temporal convolutional network}
    \acrodef{E2E}   {end-to-end}
    \acrodef{ED}    {epoch duration}
    \acrodef{ER}    {early reflection}
    \acrodef{ESTOI} {extended short-time objective intelligibility}
    \acrodef{FLOPS} {floating operations per second}
    \acrodef{FT}    {fine-tuning}
    \acrodef{FTE}   {fine-tuning epoch}
    \acrodef{FFT}   {fast Fourier transform}
    \acrodef{GFLOPS} {giga floating operations per second}
    \acrodef{gLN}   {global layer normalization}
    \acrodef{GN}    {group normalization}
    \acrodef{GLU}   {gated linear unit}
    \acrodef{GRU}   {gated recurrent unit}
    \acrodef{GPIT}  {guided permutation invariant training}
    \acrodef{GPU}   {graphics processing unit}
    \acrodef{ICA}   {independent component analysis}
    \acrodef{IN}    {instance normalization}
    \acrodef{IPD}   {inter-phase difference}
    \acrodef{IR}    {impulse response}
    \acrodef{L} {large}
    \acrodef{LD}    {logit difference}
    \acrodef{LGRU}  {light gated recurrent unit}
    \acrodef{LM}    {language model}
    \acrodef{LN}    {layer normalization}
    \acrodef{LR}    {late reflection}
    \acrodef{LSTM}  {long short term memory}
    \acrodef{LTI}   {linear time-invariant}
    \acrodef{LUFS}   {loudness units relative to full scale}
    \acrodef{M} {medium}
    \acrodef{MACs}  {mutiply-accumulate operations}
    \acrodef{MHA}   {multihead attention}
    \acrodef{MHSA}  {multihead self-attention}
    \acrodef{MHCA}  {multihead cross-attention}
    \acrodef{MIMO}  {multiple-input multiple-output}
    \acrodef{ML}    {machine learning}
    \acrodef{MLP}   {multi-layer perceptron}
    \acrodef{MOS}   {mean opinion score}
    \acrodef{MPGT}  {multi-phase gammatone}
    \acrodef{MR}    {mask refinement}
    \acrodef{MRD}   {mask refinement decoder}
    \acrodef{MSE}   {mean square error}
    \acrodef{MVDR}  {minimum-variance distortionless response}
    \acrodef{MT}    {machine translation}
    \acrodef{NN}    {neural network}
    \acrodef{NMF}   {non-negative matrix factorization}
    \acrodef{NSM}   {noisy speech mixture}
    \acrodef{NRSM}  {noisy reverberant speech mixture}
    \acrodef{OOM}   {out of memory}
    \acrodef{ORC-WER}{optimal reference combination word error rate}
    \acrodef{PCA}   {principal component analysis}
    \acrodef{PE}    {positional encoding}
    \acrodef{PESQ}  {perceptual evaluation of speech quality}
    \acrodef{PIT}   {permutation invariant training}
    \acrodef{PM}    {post-masking}
    \acrodef{PMD}   {post-masking decoder}
    \acrodef{PReLU} {parametric rectified linear unit}
    \acrodef{P-Conv}{\texorpdfstring{pointwise convolution}{P-Conv}}
    \acrodef{QDPN}  {quasi-dual-path network}
    \acrodef{QRNN}  {quasi-recurrent neural network}
    \acrodef{ReLU}  {rectified linear unit}
    \acrodef{RF}    {receptive field}
    \acrodef{RIR}   {room impulse response}
    \acrodef{RNN}   {recurrent neural network}
    \acrodef{RNN-T}   {recurrent neural network transducer}
    \acrodef{RSM}   {reverberant speech mixture}
    \acrodef{RTF}   {real time factor}
    \acrodef{RQ}    {research question}
    \acrodef{S} {small}
    \acrodef{SA}    {self-attention}
    \acrodef{SAD}   {self-attention decoder}
    \acrodef{SAE}   {self-attention encoder}
    \acrodef{SC}    {skip connection}
    \acrodef{SE}    {squeeze-and-excite}
    \acrodef{SepFormer}{separation Transformer}
    \acrodef{SISDR} {scale-invariant signal-to-distortion ratio}
    \acrodef{SRMR}  {speech-to-reverberation modulation energy ratio}
    \acrodef{SDR}   {signal-to-distortion ratio}
    \acrodef{SiLU}  {sigmoid linear unit}
    \acrodef{SISO}  {single-input single-output}
    \acrodef{SNMF}   {sparse non-negative matrix factorization}
    \acrodef{SNR}   {signal-to-noise ratio}
    \acrodef{SOT}   {serialized output training}
    \acrodef{SOTA}  {state-of-the-art}
    \acrodef{SP}    {signal processing}
    \acrodef{SSR}   {speech-to-speech ratio}
    \acrodef{SSSR}  {self-supervised speech representation}
    \acrodef{STFT}  {short-time Fourier transform}
    \acrodef{STOI}  {short-time objective intelligibility}
    \acrodef{SVCCA} {singular vector canonical correlation analysis}
    \acrodef{SW}    {shared weights}
    \acrodef{TasNet}{time-domain audio separation network}
    \acrodef{TC}    {time-complexity}
    \acrodef{TD-Conformer}{time domain Conformer}
    \acrodef{TCN}   {temporal convolutional network}
    \acrodef{TF}    {time-frequency}
    \acrodef{TPU}   {tensor processing unit}
    \acrodef{TSE}   {target speaker extraction}
    \acrodef{TSL}   {training signal length}
    \acrodef{uPIT}  {utterance-level permutation invariant training}
    \acrodef{UPGMA} {unweighted pair group method with arithmetic mean}
    \acrodef{WD-Conv}{weighted multi-dilation depthwise convolution}
    \acrodef{WD-TCN}{weighted multi-dilation temporal convolutional network}
    \acrodef{WE}    {Whisper encoder}
    \acrodef{WER}   {word error rate}
    \acrodef{WPE}   {weighted prediction error}
    \acrodef{XL} {extra-large}

%% file: tables/main_results.tex
\begin{table*}[!tb]
\caption{ASR and speech enhancement performance comparison of the \ac{AE} loss in (\ref{eq:ae_loss}) to baseline SISDR loss models as well as oracle signal $s_c[i]$ and mixture signal $x[i]$ after a set number of \acp{ATE}. $\Delta$ indicates improvement over the mixture signal. {Bold} font indicates the best result for each metric. Arrows indicate performance increase.}
\centering
\label{tab:wav2vec2}
\footnotesize
\begin{tabular}{|c|c|c|cc|cc|c|c|c|c|}
\hline
\rowcolor[HTML]{C0C0C0} 
\textbf{ASR Input} &
  \textbf{Loss} & \textbf{ATEs} &
  \textbf{CP-WER} $\downarrow$ &
  \textbf{$\Delta$ $\uparrow$} &
  \textbf{ORC-WER}  $\downarrow$&
  \textbf{$\Delta$ $\uparrow$} &
  \textbf{SISDR} $\uparrow$&
  \textbf{PESQ} $\uparrow$&
  \textbf{STOI} $\uparrow$&
  \textbf{SRMR} $\uparrow$\\ \hline
Oracles $s_c$         & -                            & - & 7.7 & - & 7.7  & - & - & - & - & - \\
Mixture  $x$          & -                            & - & 85.7 & - & 81.8 & - & -7.2 dB & 1.05 & 27.1 & 2.99\\ \hline
Estimates  $\hat{s}_c$ & $\mathcal{L}_\mathrm{SISDR}$ & - & 44.2 & 41.5 & 44.2 & 37.6 & 5.5 dB & 1.41 & 69.3 & 7.14\\ \hline
Estimates  $\hat{s}_c$ & $\mathcal{L}_\mathrm{SISDR}$ &   30 & 43.5 & 42.3 & 43.4 & 38.4 & \textbf{5.6} dB & 1.44 & 71.9 & 7.21\\ \hline
Estimates  $\hat{s}_c$ & $\mathcal{L}_\mathrm{AE}$ (proposed)   &   30  & \textbf{37.1} & \textbf{48.7} & \textbf{37.1}  & \textbf{44.8} & 5.1 dB & \textbf{1.54} & \textbf{75.0} & \textbf{7.27}\\
\hline
\end{tabular}%
\end{table*}

%% file: tables/whisper_results.tex
\begin{table}[!th]
\caption{\Ac{ASR} performance for the re-evaluation of models in \autoref{tab:wav2vec2} using the large Whisper model \cite{whisper}. Bold indicates the best-performing trained model. $\Delta$ indicates improvement over the mixture signal.}
\footnotesize
\centering
\label{tab:whisper}
\begin{tabular}{|c|c|c|cc|}
\hline
\rowcolor[HTML]{C0C0C0} 
\textbf{ASR Input} &
  \textbf{Loss} & \textbf{{ATE}s} &
  \textbf{CP-WER} $\downarrow$ &
  \textbf{ORC-WER} $\downarrow$
  \\ \hline
Oracles $s_c$         & -                            & - & 10.9 & 10.9 \\
Mixture  $x$          & -                            & - & 63.3 & 60.0\\ \hline
Estimates  $\hat{s}_c$ & $\mathcal{L}_\mathrm{SISDR}$ & - & 29.5 & 29.4 \\ \hline
Estimates  $\hat{s}_c$ & $\mathcal{L}_\mathrm{SISDR}$ & 30 & 29.3 & 29.2 \\ \hline
Estimates  $\hat{s}_c$ & $\mathcal{L}_\mathrm{AE}$    & 30  & \textbf{26.8} & \textbf{26.7} \\
\hline
\end{tabular}%
\end{table}

%% file: tables/TSL_table.tex
\begin{table}[!th]
\caption{Comparison of the impact of the \ac{TSL} limit on \ac{ASR} performance of models fine-tuned using the proposed \ac{AE} loss over 30 \acp{ATE}. Best results are shown in bold.}
\label{tab:tsllimits}
\footnotesize
\centering
\begin{tabular}{|c|c|c|c|}
\hline
\rowcolor[HTML]{C0C0C0} 
\textbf{TSL limit (s)} & \textbf{CP-WER} $\downarrow$ & \textbf{ORC-WER} $\downarrow$ & \textbf{SISDR} $\uparrow$\\ \hline
2.0                    &       38.3          &         38.4         &         \textbf{5.0 }      \\ \hline
4.0                    &       37.1          &         37.1         &        5.1        \\ \hline
8.0                    &       \textbf{36.3}          &       \textbf{ 36.2 }         &         5.2       \\ \hline
\end{tabular}%
\end{table}